# A small world of citations?
# The influence of collaboration networks on citation practices

Matthew L. Wallace[1], Vincent Larivière[2] and Yves Gingras[3]

*[1] matt.l.wallace@gmail.com*
Centre interuniversitaire de recherche sur la science et la technologie (CIRST), Université du Québec à Montréal, Case Postale 8888, Succursale Centre-Ville, Montréal, Québec, H3C 3P8 (Canada)

*[2] lariviere.vincent@uqam.ca*
École de bibliothéconomie et des sciences de l'information, Université de Montréal, C.P. 6128, Succ. Centre-Ville, Montréal, QC. H3C 3J7, Canada  and
Observatoire des sciences et des technologies (OST), Centre interuniversitaire de recherche sur la science et la technologie (CIRST), Université du Québec à Montréal, Case Postale 8888, Succursale Centre-Ville, Montréal, Québec, H3C 3P8 (Canada) and
Cyberinfrastructure for Network Science Center, School of Library and Information Science, Indiana University, 10th Street & Jordan Avenue, Wells Library, Bloomington, Indiana, 47405 (United States)

*[3] gingras.yves@uqam.ca*
Observatoire des sciences et des technologies (OST), Centre interuniversitaire de recherche sur la science et la technologie (CIRST), Université du Québec à Montréal, Case Postale 8888, Succursale Centre-Ville, Montréal, Québec, H3C 3P8 (Canada)

**Abstract**

This paper examines the proximity of authors to those they cite using degrees of separation in a co-author network, essentially using collaboration networks to expand on the notion of self-citations. While the proportion of direct self-citations (including co-authors of both citing and cited papers) is relatively constant in time and across specialties in the natural sciences (10% of citations) and the social sciences (20%), the same cannot be said for citations to authors who are members of the co-author network. Differences between fields and trends over time lie not only in the degree of co-authorship which defines the large-scale topology of the collaboration network, but also in the referencing practices within a given discipline, computed by defining a propensity to cite at a given distance within the collaboration network. Overall, there is little tendency to cite those nearby in the collaboration network, excluding direct self-citations. By analyzing these social references, we characterize the social capital of local collaboration networks in terms of the knowledge production within scientific fields. These results have implications for the long-standing debate over biases common to most types of citation analysis, and for understanding citation practices across scientific disciplines over the past 50 years. In addition, our findings have important practical implications for the availability of 'arm's length' expert reviewers of grant applications and manuscripts.

## 1. Introduction

Scientific collaborations and citation practices have been an important focus of interest among sociologists of science, seeking to provide insight into science as an inherently social and team-based endeavour. While co-authorship networks are relevant for understanding the network structure of scientific fields (Jones *et al.*, 2008), citation practices are central to the distribution of symbolic capital and its accumulation by scientists and provide insight into the hierarchies within a field and among fields (Whitley, 1984). Though everything suggests that some relationship must exist between co-authorship and citation practices, these two elements are generally treated separately and few papers have addressed that question. White *et al.* (2003) combined, for a small group of researchers, bibliometrics with survey data to see whether citations were influenced by the social structure of the group. Introducing the notion of 'inter-citation' as a measure of citations between members of a given group, they aimed to correlate citations with social, socio-cognitive and intellectual ties. Their conclusions, based on only 16 individuals, are nuanced: there is some correlation, as one might expect, between collaboration and citation patterns but, overall, there is no strong or reliable link between social ties and citations (see also Johnson and Oppenheim, 2007 and Rowlands, 1999 for related studies).

Most recently, Ding (2011) used a relatively large dataset to systematically explore a similar issue, that of the proximity of authors (in terms of collaboration) within and between various research topics in the field of information retrieval. A similar treatment is applied to citation networks, which reveals much about the intellectual cohesiveness of certain sub-fields, but does not reveal the degree to which social networks affect the referencing system. Recent studies such as that of Roth and Quintet (2010) have successfully combined social and semantic networks as a means to understand the production of knowledge within 'epistemic communities'.

Using a very large dataset (over 2,6M papers and 50M references) over more than 50 years, this paper combines and expands on previous methods for analyzing co-author networks and for measuring self-citations. It poses the all-important question of whether the social network of researchers has an impact on the selection of references found in a given article. In contrast to White *et al.* (2003), we restrict ourselves to co-authorship as an indicator of their social proximity. Collaboration networks can be considered as a subset of the complete social network of a scientist. Though one usually knows more scientists than the ones with whom one writes scientific papers, it seems natural to consider co-authors as part of that social network even in the case where no face-to-face interactions has occured. Moreover, the ties with co-authors are probably stronger than with non co-authors and thus their effect on citation should be larger than with non co-authors even if the latter are part of the larger social network of the citing scientist.

In this paper we thus analyze the references of each article in terms of four levels of social proximity, defined as co-authors or co-authors of co-authors in analogy with the concept of Erdös numbers (see Methods section below). This study performs the first large-scale investigation of the connections between citation patterns and collaboration networks. In order to distinguish between a variety of citation practices within the natural and medical sciences (NMS) and social sciences and humanities (SSH), eight specialities were chosen for detailed analysis. After a detailed description of the methods and database used (Part 2) we present the results (Parts 3) and discussion (Part 4) followed by a conclusion highlighting some implications of our findings (Part 5).

## 2. Methods

The data for this analysis comes from Thomson Scientific's Web of Science, which include the Science Citation Index Expanded (SCIE), Social Science Citation Index (SSCI), and Arts and Humanities Citation Index (AHCI) for the 1945–2008 period. Data is presented for 8 specialities (5 from the Natural and medical sciences, 3 from the Social sciences and humanities) based on the U.S. National Science Foundation (NSF) field classification[1]: astronomy and astrophysics, atmospheric science and meteorology, biochemistry and molecular biology, economics, history, neurology and neurosurgery, organic chemistry and sociology. Only research articles, notes and reviews are included in the set.

There are two main methodological challenges to measuring how socially close citing authors are to those they cite. First, we need to conduct a large-scale analysis to measure the social proximity of cited authors to citing authors for many different scientific disciplines in order to capture the diversity of practices. Second, the analysis needs to be focused on the individual authors, in order to gain insight into their referencing practices and individual social networks.

---

[1] http://www.nsf.gov/statistics/seind06/c5/c5s3.htm#sb1

In order to investigate the citation practices of a given scientific specialty in relation to its co-authorship network, we form a set of references $R_k$ contained in the set of papers $S_k$ published in a given year $k$ within a given specialty. Given that Thomson Reuters' Web of Science only indexes the names of co-authors of cited papers that are also *source* items, we restrict this set of references to those who can be identified within the database as source items (Snyder & Bonzi,1998), and which were published within the previous 10 years. We generate a list of authors $a_k$ having contributed to each article $s_k \in S_k$, yielding a total set of authors $A_k$ for the specialty as a whole.

Similarly, we generate a second set of authors $a'_k$ (and $A'_k$ for the entire specialty) who collaborated within 2 years[2] of the publication year $k$ with authors in $a_k$ (restricted to the specialty in question in order to limit false positives due to the presence of homonyms[3]). Thus, $A'_k$ constitutes the unweighted and undirected co-author network. Finally we generate a third group of authors $a''_k$ who collaborated with $a'_k$ during the same time period. It should be noted that $a''_k$ excludes all authors contained in $a'_k$, so in general, for networks which are relatively sparse, or which contain few numbers of co-authors, $n(A''_k) < n(A'_k)$ (while the opposite is true for cases when collaboration rates are high).

For each source article $s \in S$, we examine its set of references and classify them in the following way:
A) If any of the authors of the referenced paper is contained in $a_k$, then this is a **self-citation**;
B) If any of the authors of the referenced paper is contained in $a'_k$, then this is a **level-1 co-author citation**;
C) If any of the authors of the referenced paper is contained in $a''_k$, then this is a **level-2 co-author citation**;
D) If none of the authors in the referenced paper are contained in $a_k$, $a'_k$, or $a''_k$, then this is called **distant citation**.

These categories are defined as mutually exclusive: if a referenced paper can be placed in more than one category, then it is assigned the one closest to a self-citation. References falling into categories A are obvious self-citations while those falling under B and C will hereafter be referred to as co-authorship citations. Level 1 and 2 co-author citations can also be seen as a measure of social proximity of the co-authors, with level 1 being closer to the author than level 2.

Many will recognize these levels as the beginning of the Erdös number or degrees of separation game (*cf.* Batagelj & Mrvar, 2000), applied to each author individually, and his referenced authors as the 'object' of the game. From a sociology of science perspective, it is not necessary to continue past the second 'level' (Erdös number of 2), since we can consider that there is no longer any social connection between the authors within a given specialty. In addition, given the number of authors and references being considered, the data mining procedure is both expensive in CPU time and memory usage. Figure 1 provides a visual representation of this algorithm.

---

[2] Because of this $\pm 2$ year interval, the data presented is for the 1947-2006 period, though data is collected for the 1945-2008 period.
[3] Given that no treatment were performed in order to distinguish authors having the same surname and initial(s) (homonyms), our data is likely to overestimate self-citations for specialties with high levels of co-authorship. Nevertheless, based on small samples, we estimate the number of homonyms at less than 5% of total positives.

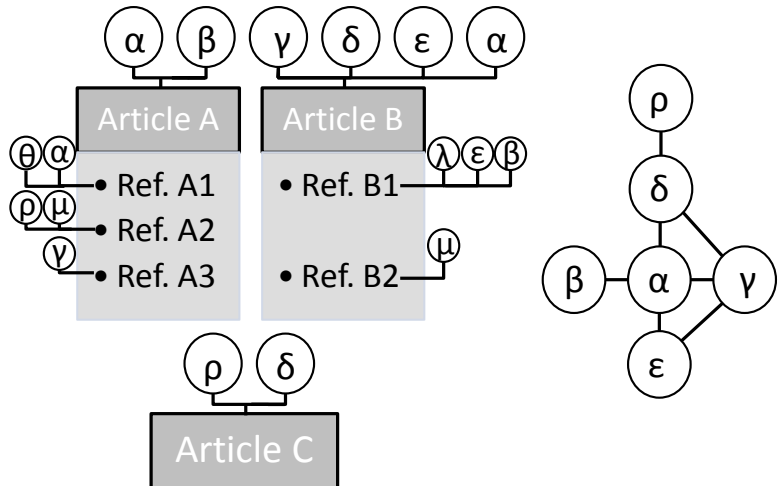

**Figure 1: An illustrative representation of the algorithm. Left: a set of three articles and 5 references therein. Right: The corresponding co-author network. Article A, for example is written by two authors (α and β) and contains three references (whose authors are also denoted by Greek letters). Based on our classification scheme of co-authorship citations, references A1 and B1 are** *self-references*, **A2 is a** *level-2 co-author citation* **(since α collaborated with δ who collaborated with ρ), A3 is a a level-1 co-author citation and B2 is a** *distant citation*.

Finally, it should be noted that while the citing source items and authors are restricted to a given specialty, the items they cite are not. One would expect that the specialty in question covers the majority of 'peers' cited, but such a limit, while defining a 'closed' system, would possibly introduce an artefact, particularly for more interdisciplinary specialties such as biochemistry (see Figure 2C). However, in such cases, we have checked that the results are similar, whether or not we restrict the specialty of the reference items.

### 3. Results

*A) Macroscopic properties of scientific specialties*

Based on the dataset described in the previous section, we first compute a few basic macroscopic variables which allow us to characterize the growth and structure of the chosen fields. The number of papers, the rate of co-authorship and the authors' productivity (Figures 2A, 2C and 2D) provide insight into the social structure and size of the discipline, while the number of references, proportion of *intra*-disciplinary references (e.g., economics to economics) and age of references (Figure 2B, 2E and 2F) provide information on the different citation practices. More generally, these data provide a benchmark for characterizing the production of scientific knowledge in various fields, and thus understanding our data on the proximity of citing and cited authors.

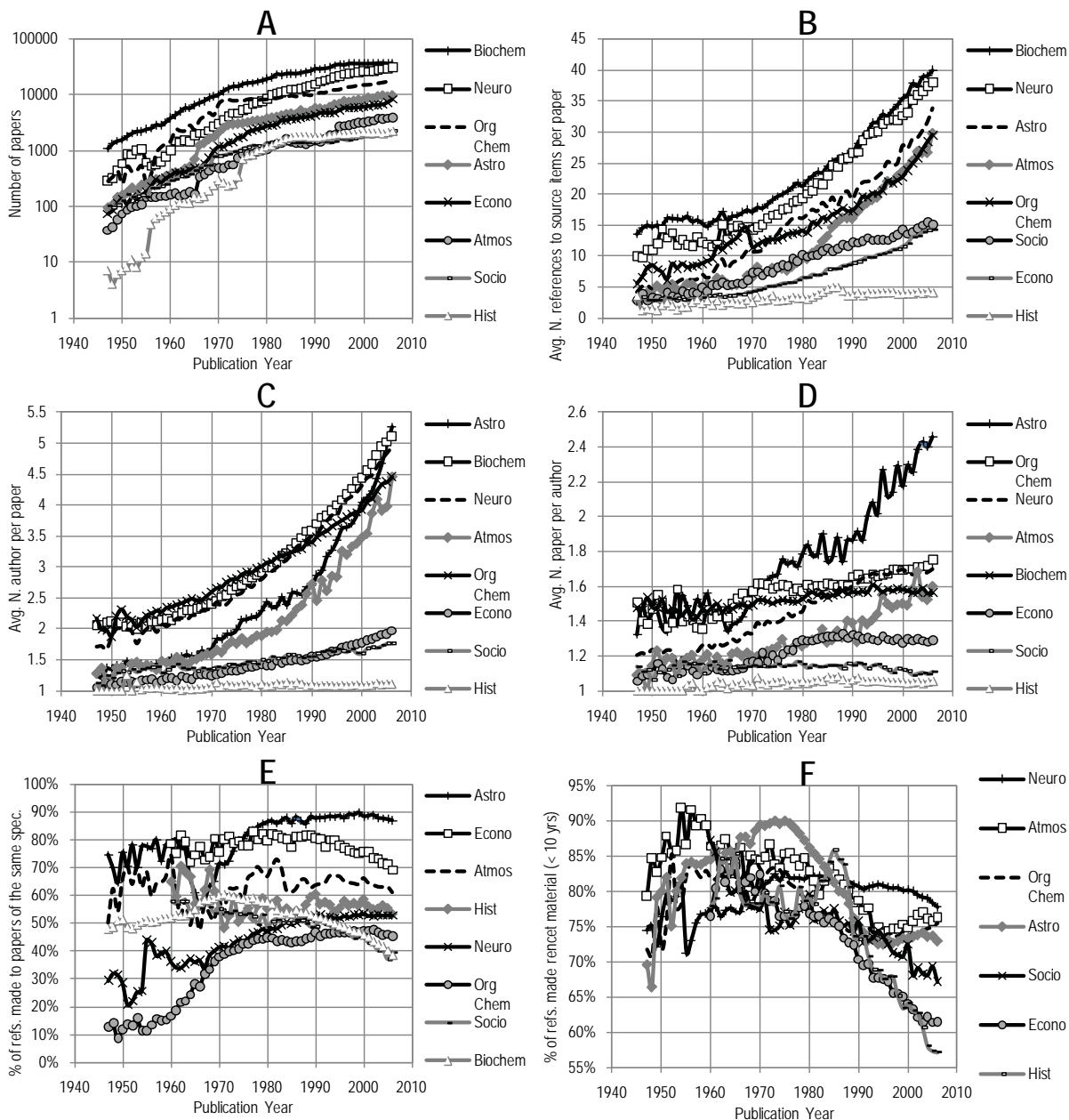

**Figure 2. For each of the chosen specialties, A) number of papers, B) average number of references per paper, C) average number of authors per paper, D) average number of papers written by each author, E) percentage of identified references within the same specialty, and F) percentage of identified references defined as 'recent' (less than 10 years older than the source item).**

*B) Empirical evidence for co-authorship citation and macroscopic properties of scientific specialties*

Figure 3 shows the distribution of citations across several specialties in the NMS (panels A-E) and the SSH (panels F-H). As one might expect, the proximity of references in each of the disciplines varies a great deal. Within the natural sciences, one immediately notices a major difference in the co-authorship proximity of references between, on the one hand, astrophysics/astronomy and atmospheric science and meteorology, and the rest of the specialties on the other hand. Aside from organic chemistry, all specialties show a clear decrease in the percentage of references made to distant authors with whom, according to our definition, they have no close connection. Furthermore, while it is clear that the size of the specialties (Figure 2A), the number of co-authors per paper (Figure 2B), and the proportion of 'intra-specialty' references (Figure 2E) have a clear impact on the proximity of references (as one might expect), none of these macroscopic quantities can singlehandedly explain the trends observed in Figure 3. In addition, there is no strong correlation between the tendency to cite recent literature (Figure 2F) and the proportion of that literature that is socially proximate.

Direct self-citation, however, is relatively constant both across fields and over time, hovering around 20% in NMS specialties and 10% in the SSH. Note that studies of various disciplines have found rates of self-citations among references varying between 10% and 36%, with strong variations between specialties (Tagliacozzo, 1977, Lawani, 1982, MacRoberts & MacRoberts, 1989), and much lower percentages in SSH such as sociology and economics (Bonzi and Snyder, 1990). Our data mostly agree with these numbers, although none of the specialties analyzed here obtain a number as high as 36%. It has also been found (Glänzel et al., 2006) that 1) self-citations are generally younger and have a shorter half-life than foreign citations, 2) self-citations stabilize in a period of 3-4 years after publication and 3) the percentage of self-citations only slightly increases with the number of co-authors.

The difference between the NMS and SSH is substantial, and dwarf the differences among SSH specialties shown in Figure 3. We find that there is no such thing as co-authorship citations within the three SSH fields studied. This is primarily due to the fact that co-authorship is less frequent in these disciplines and that, as a consequence, researchers have less co-authors in their social network to choose from, a clear limitation in the way we define our social network. For this reason, most of our analysis focuses on the NMS.

For NMS disciplines, we also show the corresponding distribution of references when we limit the set $S_k$ for each year to papers with 5 co-authors or less (gray dashed lines in Figure 3). While arbitrary, this immediately gives us a sense of the extent to which disciplines such as astrophysics and astronomy cite a larger proportion of papers authored by their recent co-authors due to the large number of papers with a large number of co-authors. Furthermore, it is more likely that authors of papers with 5 authors or less actually *know* each other. For clarity, we omit from Figure 3 the number of self-references, references to level-1 co-authors and to level 2 co-authors when this restriction is imposed. Interestingly, the increase in distant citations observed is at the expense of level-1 and level-2 co-authors citations, but does not affect self-citations.

This remarkable stability in the level of self-citations—across specialties and time—distinguishes this practice from that of citing those who have been recent collaborators (not just on the particular paper in question). This suggests that there might be cross-disciplinary norms regarding this practice in science. It must be noted that this does not imply a degree of conformity *within* the specialty— comparison of the distributions of self-citations would be more revealing in this respect. However, the stability of the average is important in understanding that this practice does not depend much on the number of co-authors or the citation practices of the discipline, but is a widespread and stable practice in all disciplines. For this reason and due to the increasing importance of research groups as a dominant unit for understanding scientific work, it is important to analyse co-author citations, which reflect the social proximity of citing and cited authors. By contrast, focusing on distant citations sheds light on the communication structure of scientific specialties by pointing at possibly independent sub-groups who are not in direct contact with each other through co-authorship links but are nonetheless cited. This approach can complement a more micro-level analysis of the reasons scientists invoke for citing papers (Milard 2010).

It should be noted that the effect of having a large number of collaborators per paper amplifies the proportion level-1 and level 2 proximate citations. Our findings clearly show that recent increases in the proximity of citing and cited authors are, in part, due to an increase in the size of collaborations. This is the case in astrophysics and astronomy, for instance. Co-authorship practices in fields such as astrophysics or particle physics often reflect the use of certain instruments or of a willingness to acknowledge the contributions of a wider range of individuals in the division of labour, beyond the writing of the article itself (Biagioli, 2003). In this sense, there is inevitably a sociological basis to this combinatorial effect.

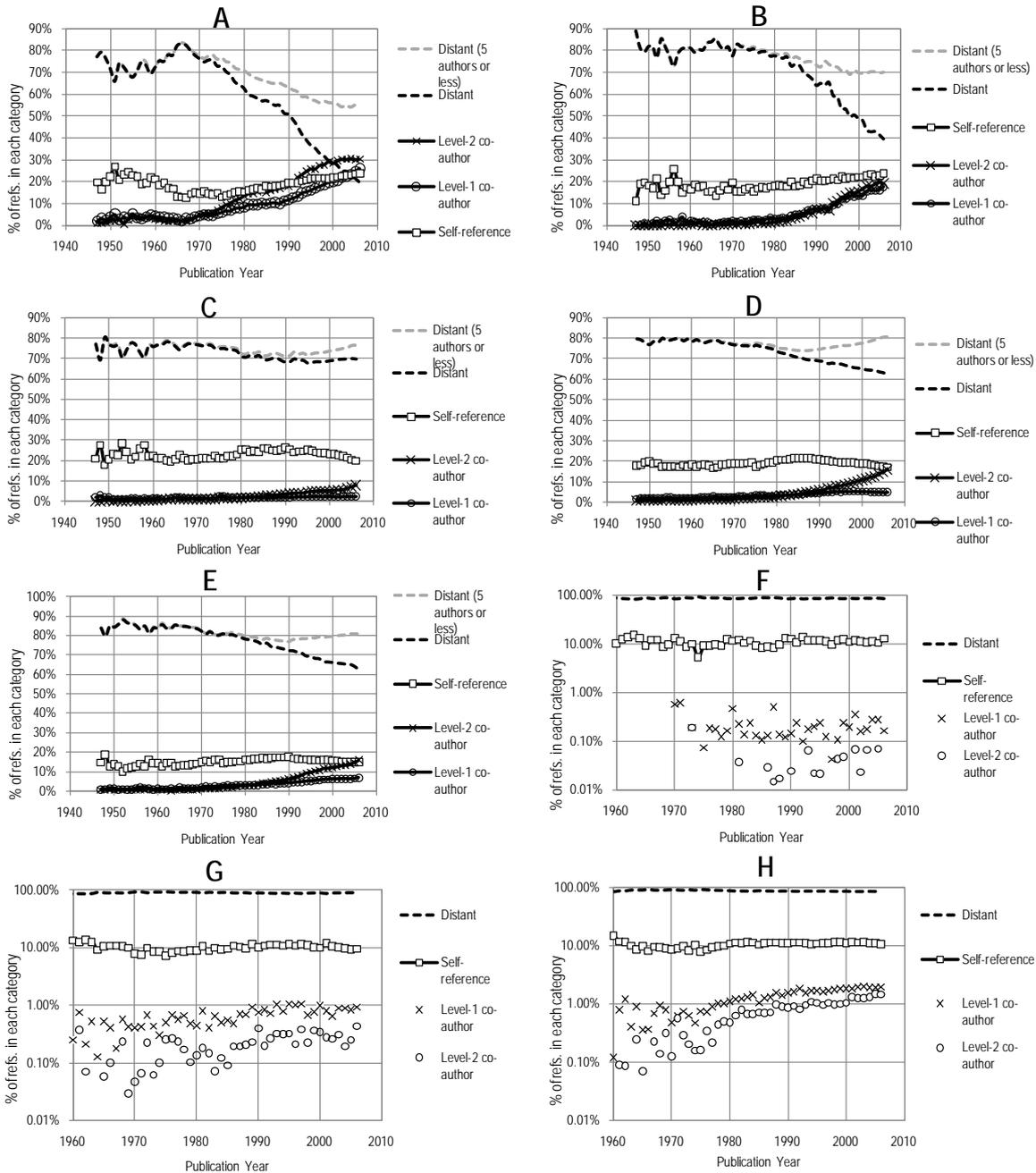

**Figure 3.** The distribution of references made in five natural science and three social science specialties: A) astrophysics and astronomy, B) atmospheric science and meteorology, C) organic chemistry, D) biochemistry and molecular biology, E) neurology and neurosurgery, F) history, G) sociology and H) economics. The last three (F, G, H) are shown on a logarithmic scale for clarity (which explains why 'distant' references seem to be close to 100%). For the NMS, we compute the same distribution based on a subset of source articles (and their references) that contain only 5 authors or less (dashed gray lines).

Our results also clearly show that the combinatorial effect cannot alone account for the proximity of citing and cited authors. Indeed, from a social network perspective, the co-author network is defined by more than the distribution of edges per node. In other words, it is not just about how large collaborations are, but also of what type of collaborations occur and where. We have also found that the distribution of clustering coefficients (Watts & Strogatz, 1988) is very similar in the co-author networks of five NSM scientific specialties in recent years. This index essentially measures the concentration of triangles within the network or to what extent collaborators of a given author also collaborate with each other. Therefore, other measures should be able to account for the local structure of the networks. Along the lines of Moody (2004), we view self-citation and 'group' citation as a means to reinforce local social networks, which has particular importance for the intellectual and social development of scientific specialties. And given that the network embodies a form of social capital of relations, and that citations tend to contribute to the symbolic capital accumulated by the individuals cited, one can see level-1 and level-2 citations as a kind of transformation of social capital into symbolic capital (Bourdieu, 1986). But this symbolic capital being a rare and contested resource, scientists who contest its value for a given scientist could identify these co-author citations with a self-citation, which tend to be perceived in a negative manner, thus annihilating its value. At the analytical level, it is thus important to distinguish self-citations from level-1 and level 2 citations, though actors could try to extend the negative connotations of the former to the latter.

*C) Topology of the networks and citation practices*

Reducing the number of co-authors to 5 or less is not sufficient to understand to what degree the number of other authors in proximity to a given author influences their choices of citations. This begs the question of how the number of level-1 and level-2 co-authors is distributed within each of the specialties. Figure 4 shows these distributions for two periods: 1960-1969 and 2000-2006. Two main observations can be made. First, variations in distributions of co-authors do not correlate highly with differences in the number of social references (Figures 3A-E), so other factors must be also at play. Second, the relatively even distribution of level-2 co-authors means that, within a given network, there will be wide variations from paper to paper in how many of these more distant co-authors are 'available' to be cited by a given author.

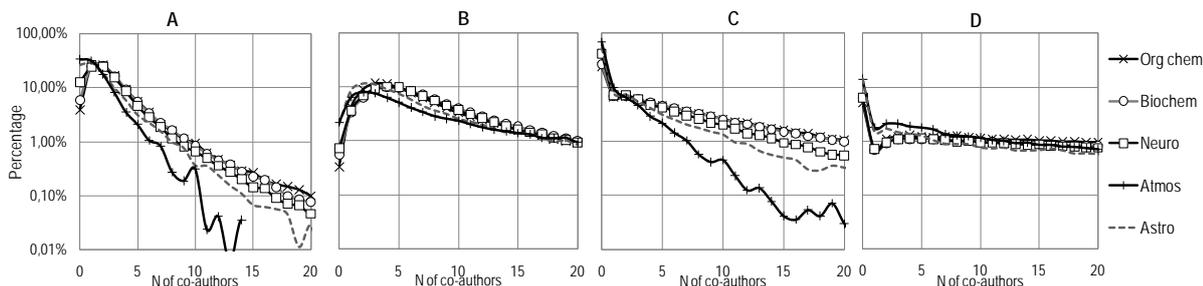

**Figure 4. Distribution of the number of A) level-1 co-authors ($a_k^{'}$) and C) level-2 co-authors ($a_k^{''}$) during the 1960-1969 period; B) level-1 co-authors ($a_k^{'}$) and D) level-2 co-authors ($a_k^{''}$) during the 2000-2006 period.**

The broadening of a distribution of co-authors cannot entirely account for increases in the proximity of citations. This is verified by randomly removing source papers (up to around 15% of the network in order to maintain its general shape) until the distributions of authors per paper are almost identical in all 8 specialties[4] and using only the first author of references. Similarly, we can randomly remove papers in a given specialty such that each author in a given interval of time has only 1 paper. These procedures have the effect of diluting the network (i.e., reducing the amount of clusters) (Newman, 2001, 2004). Once again, we see no major effect on the proximity of citations. In addition, the differences observed in proximity of citations are not (or only very weakly) reflected in measures such as the distribution of k-cores or the number of cliques. Once again, it is important to emphasize that our study examines the topology in citation proximity to each individual author, and is not concerned with the overall structure of the discipline.

---

[4] In practice, it is difficult to make the different sets of source papers have exactly the same distribution of authors per paper. Our objective is to reduce the effect of skewed distributions while ensuring that the 'reduced' network retains sociological meaning.

One of the main advantages of our method for examining referencing patterns is the ability to conduct the analysis at the level of each author or paper. It is thus useful to think of each author making referencing *choices* based in part on other authors that are in proximity to him or her. More specifically, given the number of references and authors associated with a given paper, we can consider how many of the various types of co-authorship references they select, compared to the number expected randomly. We can define this quantity as the propensity $P_d$ for a given level of proximity $d$, approximated as the ratio of the number of articles found empirically to the expected number of articles to be found given a random selection of references. The latter is nothing but a binomial distribution, so for a single article in a given year $k$, the propensity of having a level-0 (author), level-1 (co-author) or level-2 (co-author of co-author) reference can be written as:

$$P_{l0,k} = \frac{n(A_k)}{\sum_{i=1}^{n(r_k)} n(a_i)n(a_k)} n_{l0}; \; P_{l1,k} = \frac{n(A'_k)}{\sum_{i=1}^{n(r_k)} n(a_i)n(a'_k)} n_{l1} \; ; \; P_{l2,k} = \frac{n(A''_k)}{\sum_{i=1}^{n(r_k)} n(a_i)n(a''_k)} n_{l2} \quad (1)$$

where $n_{l1}$, $n_{l2}$ are the number of cases empirically identified at each level, $n(a_i)$ the number of authors of the $i^{th}$ referenced paper, $n(r_k)$ the number of remaining references of the given paper. Thus, the numerator is determined by empirical 'matches' and the size of the entire network, while the denominator reflects the size of the author's network and that of the cited authors' networks. Like our data presented in Figure 3, the propensity is computed in sequence, in order of proximity, with the 'matched' references removed at each step. In other words, the level-2 propensity, for instance, is not 'skewed' by the number of level-0 or level-1 references already found for the given paper. Similarly, if there are no available authors in the level-1 or level-2 set, then the corresponding propensity is not calculated.

Table 1. Propensity for self-citation ($P_{l0}$), as well as level-1($P_{l1}$) and level-2 ($P_{l2}$) co-authors citations.

| $P_{l0}$ | Astro | Atmos | Biochem | Econo | Hist | Neuro | Org chem | Socio |
|---|---|---|---|---|---|---|---|---|
| 1956-65 | 50 | 23 | 379 | 35 | 14 | 98 | 171 | 35 |
| 1966-75 | 210 | 70 | 840 | 91 | 31 | 269 | 537 | 69 |
| 1976-85 | 308 | 139 | 1452 | 239 | 170 | 601 | 736 | 115 |
| 1986-95 | 438 | 165 | 1818 | 352 | 234 | 888 | 819 | 141 |
| $P_{l1}$ | Astro | Atmos | Biochem | Econo | Hist | Neuro | Org chem | Socio |
| 1956-65 | 20 | 6 | 76 | 19 |  | 18 | 30 | 6 |
| 1966-75 | 50 | 27 | 139 | 26 | 5 | 56 | 56 | 22 |
| 1976-85 | 59 | 41 | 168 | 50 | 24 | 93 | 58 | 18 |
| 1986-95 | 65 | 54 | 155 | 74 | 20 | 96 | 46 | 35 |
| $P_{l2}$ | Astro | Atmos | Biochem | Econo | Hist | Neuro | Org chem | Socio |
| 1956-65 | 4,0 | 1,6 | 12,5 | 0,4 |  | 2,2 | 6,4 | 0,6 |
| 1966-75 | 7,9 | 8,6 | 20,9 | 2,2 | 0,4 | 7,9 | 10,9 | 2,0 |
| 1976-85 | 8,6 | 7,2 | 21,6 | 8,8 | 0,2 | 15,6 | 11,4 | 1,8 |
| 1986-95 | 8,7 | 7,9 | 15,9 | 11,2 | 1,1 | 13,5 | 6,9 | 3,2 |

Table 1 shows the propensity (computed individually for each source paper in the 10-year period, then averaged) for citations to level-1 and level-2 co-authors for three time periods, as described above in the Methods section. In general, there is very little propensity to level-2 citations. Data for self-citations are an order of magnitude higher than for level-1 citations to co-authors, as one might expect. Furthermore, there is a general rise in propensity since the 1950s, with a slower growth rate since the 1970s.

If a relatively large field (e.g., biochemistry and molecular biology, or economics) contains many groups working on *largely independent* topics, then the propensity for self-citation, level-1 citations and level-2 citations tends to be high. It also indicates that high levels of co-authorship citations in astrophysics and astronomy, or in atmospheric science and meteorology, are largely determined by the structure of the specialties, and less by the choices of citing authors. Thus, comparing practices in different specialties (SSH included) with very different co-author network topologies shows that level-1 citations are far from random, which likely reflects the specialization of researchers and the cumulative nature of research. Interestingly, the only two specialties which, recently, tend less and less to cite socially close authors are organic chemistry and, to a lesser degree, biochemistry. This confirms the validity of the trend observed earlier in Figure 3 and might indicate either that different types of referencing practices exist within organic chemistry (e.g., there are less perfunctory references) or that the field is less intellectually fragmented and authors search out information from further afield.

The normalized distribution of values for $P_{l1}$ and $P_{l2}$ for all papers in a given year is also revealing. If we take data from 1985, for instance, we immediately see that the level of 'zero' contributions (top-left in Figures 5A and B) vary widely among disciplines. For those articles which display a non-zero propensity there in a second, non-zero local maximum somewhere between 10 and 100, according to scientific specialty. While many papers with relatively low propensity to cite their recent authors dominates in astrophysics, only a few papers with high propensity dominate in economics or, to a lesser degree, in the atmospheric sciences. The practice is more generalized in astrophysics, but dominated by a few areas with high levels of propensity in most other fields. Once again, fragmented fields such as biochemistry have the longest tails indicating a lack of uniformity in the citation practices of its members. In the case of the propensity to cite authors in one's level-2 co-author network, the same distribution is absent, and the overall lack of references of this type (compared with the 'random' case) means that there is no non-zero local maximum.

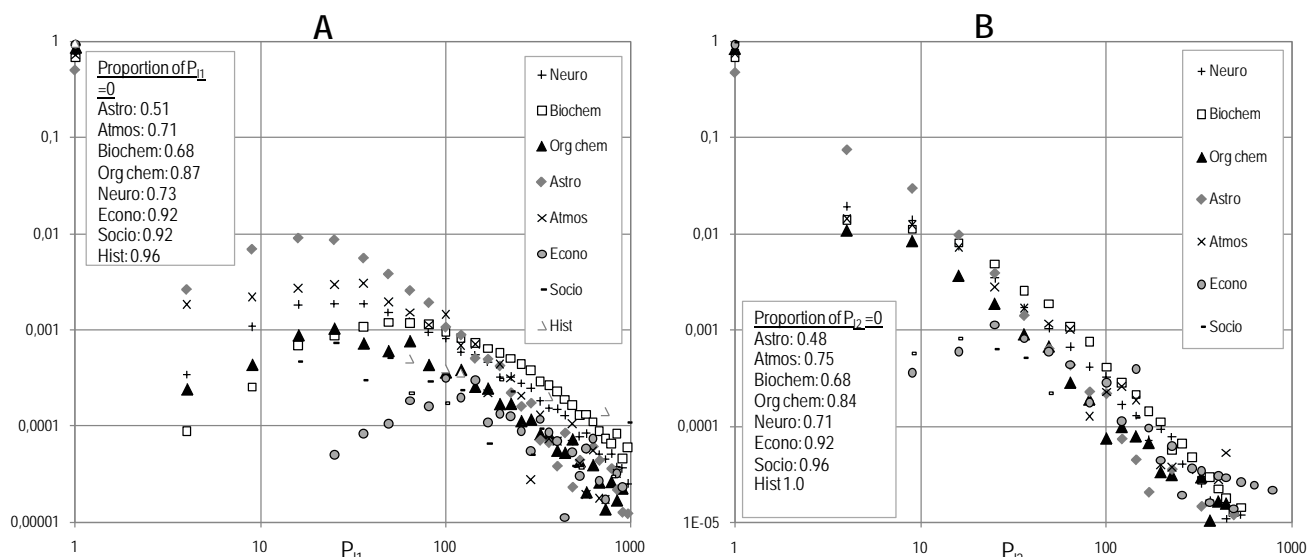

**Figure 5. Distribution of $P_{l1}$ (A) and $P_{l2}$ (B) for all papers published in 1985 in the disciplines selected. The normalized distribution is computed using a logarithmic binning scheme and the values of the y-intercepts (number of articles which don't cite any level-1 or level-2 co-authors) are listed in each figure.**

Ranking the articles in order of propensity at each level, we can also compute Spearman's rank correlation coefficient (ρ) between $P_{l0}$, $P_{l1}$ and $P_{l2}$. While there is very little correlation between $P_{l0}$ (the propensity to self-citation) and $P_{l1}$, $P_{l1}$ and $P_{l2}$ are correlated (from 0.15 in organic chemistry and 0.19 in the atmospheric sciences, to 0.40 for astrophysics). This means that authors (or, strictly speaking, papers) who tend to cite co-authors, also tend to cite co-authors of co-authors. However, given the method described above for computing the propensity, this result is not at all trivial. It implies that 'group'-citations (level-1 and level-2) are likely not similar to self-citations.

## 4. Discussion

The issue of 'compact' vs. 'fragmented' fields can only be partially explored through co-author networks. While two research groups may be entirely disconnected in terms of co-authorship proximity, they may be working on identical topics and thus still cite each other as distant citations. Since our analysis is performed at the paper level, we can only really speak of 'micro-fragmentation' at the level of small clusters. Recent studies of astrophysics, for instance, have confirmed the trends observed here of an increased reliance on a small number of journals (Abt, 2009). Some of these characteristics are also shared by the atmospheric sciences. In this case, the added fact that information is rarely sought outside the specialty and the presence of large numbers of co-authors (Figure 1) is the dominant source of the high percentage of co-authorship citations. In sociology, on the other hand, the level of co-authorship has had no effect and the field's expansion and diversification (in terms of different journals and topics covered therein) has balanced any increased propensity to cite authors who are/were also co-authors. In economics, it appears as though several of these factors may have contributed to a slight increase in co-authorship citations, though they remain extremely low. Our data clearly indicate that while the small-world phenomena observed in economics may be true due to increases in co-authorship, among other factors (Goyal *et al*., 2005), this has very little bearing on the citation practices of each author based on his or her local network.

One of the more interesting cases observed here, which remains relatively unexplored in the sociology of science literature, is that of organic chemistry, a medium-sized field with relatively high levels of co-authorship and average levels of interdisciplinarity. We have observed that co-authorship citations are low and remarkably stable, both in absolute (Figure 3) and relative terms (Table 1), and that self-citation may be declining in recent years. This would indicate that 1) perfunctory referencing is on the decline and 2) authors see decreasing value in citing their local co-author network. In fact, data from Figure 4 points to the fact that only a few organic chemists cite their collaboration network heavily, with the majority does not cite it at all.

We can also interpret these findings as measures of the value of the social networks in citation practices. In other words, to what degree is the social network of an author a determining factor in the process of producing and disseminating knowledge through publications? Within the science system, publications are, in effect, the primary means of establishing scientific authority among peers. As Bourdieu (1975, p. 24) put it, "claims to legitimacy draw their legitimacy from the relative strength of the groups whose interest they seek to express". Hyperauthorship in certain fields such as astrophysics may mean that some co-authorship citations are less perfunctory and more 'cognitive' in nature. The social network may thus 'over-determine' the citation practice. High levels of *propensity*, however, imply a need for researchers to rely on the value of their local network either for social (e.g., there is insufficient contact with other groups) or cognitive (e.g., the field is too fragmented) reasons.

In the context of a broader understanding of trends in the structure and practices of the various NMS and SSH specialty areas, our overall analysis points to the presence of more close-knit research groups in many fields, an increase of citations made to co-authors, and an increased fragmentation of research topics and groups. Recent work regarding the decline of uncitedness (Wallace *et al*., 2009) and strong evidence that scholarship is becoming less and less concentrated (Larivière *et al*., 2009) point to the fact that scholarship is not narrowing within science in general, although our data shows a correlation between fields' high levels of co-author referencing and high levels of intra-specialty citations (Figures 2 and 3).

## 5. Conclusion

Our paper expanded on the notion of self-citation to analyse the relationship between co-authorship and citation in many disciplines of the NMS and SSH, using vast quantities of data and a new algorithm. It shows that there is no single key to understanding why authors of a given specialty may cite authors with whom they, or their co-authors, have previously published. The more drastic differences among fields or over time are due to variations in levels of co-authorship, but more subtle changes are linked to the reliance of authors on their local network (and the shape of these local networks). This, in turn, is likely linked to the fragmentation of a given specialty on a small scale and the degree of intra-specialty referencing (to what degree does scholarly work build on a closed set of journals). More specifically, we have shown that:

- The gap in co-author citations between the social sciences and natural sciences remains very large, due to the different levels of co-authorship and citation practices of the actors.
- Self-citation is constant in time and across specialties of the natural sciences and the social sciences (where it is much lower), and is not dependent on the size of networks or the citation practices of actors.
- The propensity to cite co-authors and co-authors of co-authors varies widely among fields (when compared to what would be expected given the number of references per paper and size of the network). Within each field (particularly in the social sciences and less in astronomy and astrophysics), the distribution of these propensities also reveals a great deal of heterogeneity in the set of papers.
- Authors who tend to cite collaborators will also tend to cite collaborators of collaborators.

Our results can thus help temper and qualify some of the recurring concerns related to the manipulation of research evaluation data through 'citation cartels', for instance, for which large-scale empirical data has been lacking (MacRoberts and MacRoberts, 1989; Franck, 1999; Phelan, 1999). More generally, proximate referencing is often regarded as a perversion of the citation process, and seen as evidence that a field is too inward-looking or controlled by a small number of authors. A recent article by Bras-Amorós (2011) highlights this point, by analyzing the citation 'distance' as an impediment to their quality. Our analysis suggests that this is not necessarily the case. Indeed, co-authorship itself can have many meanings, not only in terms of division of labour, but also as a means of establishing a hierarchy within a field. The formation of large groups with high levels of social capital, using each other's work and collaborating more or less frequently, does not necessarily imply 'citation cartels' or nepotism. However, it is true that the high socio-cognitive 'compactness' of fields such as astrophysics and astronomy, or meteorology and the atmospheric sciences, might pose certain problems. For instance, it can be more difficult to locate 'unbiased', arm's length reviewers of papers and grants.